\documentclass[aps,10pt,final,titlepage,oneside,twocolumn,nobibnotes,nofootinbib,superscriptaddress,noshowpacs,centertags]{revtex4}
\usepackage{amssymb,amsmath}
\usepackage{eufrak}
\usepackage{mathtext}
\usepackage[cp1251]{inputenc}
\usepackage[T2A]{fontenc}
\usepackage[russian,english]{babel}
\usepackage{bm}
\usepackage{graphicx}
\usepackage{calc}
\usepackage{ifthen}
\usepackage{caption2}
\usepackage{epstopdf}
\usepackage{multirow}
\begin{document}
\selectlanguage{english}

\title{Aging effects in critical behavior of Heisenberg anisotropic ultrathin films}
%

\author{Maria Shlyakhtich$^{2,}$}
\affiliation{Siberian Federal University, 79 Svobodny Av., Krasnoyarsk, 660041, Russia }

\author{Pavel Prudnikov}
\affiliation{Dostoevsky Omsk State University, Mira prospekt 55-A, Omsk, 644077, Russia}

E-mail: mshlyakhtich@sfu-kras.ru

\begin{abstract}

We present the results of Monte-Carlo studies of the non-equilibrium properties of ferromagnetic Heisenberg
films. Aging effects were observed in non-equilibrium critical behavior. The calculations were carried out for both
high-temperature and low-temperature initial states. The characteristic correlation time, which diverges at the
transition temperature in the thermodynamic limit, was obtained as a function of system size and waiting time.

\end{abstract}

\maketitle

\section{INTRODUCTION}

A large number of experimental works \cite{bib:01, bib:02, bib:03} are devoted to the study of various properties of ultrathin films, including magnetic ones. Interest in such objects is very high due to the wide range of practical applications of these systems. Due to the strong influence of the shape and crystallographic anisotropy of the substrate, magnetic ordering in ultrathin ferromagnetic films is very difficult. In this regard, the theoretical calculations of spin models and the development of computer simulation methods are important for the rationalization and management of new experiments.

A large number of phenomena appear in statistical systems with slow dynamics. These phenomena include: a sharp slowdown in relaxation processes, memory effects, aging effects, etc. In view of this, systems with slow dynamics have recently attracted great theoretical and experimental interest \cite{bib:04,bib:05,bib:06,bib:07}.

After a long time, a system with slow dynamics does not reach equilibrium even after a small perturbation. In connection with this, its dynamics is not invariant to either time transfer or time reversal, as is usually the case in thermal equilibrium. The effects of aging appear during this endless relaxation. Thus, two-time quantities, such as the response and correlation functions, depend on two times: the waiting time $t_w$ and the observation time $t - t_w$ $(t > t_w)$, and their damping as a function of $t$ is slower at large $t_w$. In contrast to one-time quantities (for example, the order parameter) converging to asymptotic values in the limit of large times, two-time quantities are clearly characterized by signs of aging.

\section{MODEL DESCRIPTION}\label{sec_1}

In this work we study the ferromagnetic thin film with Heisenberg hamiltonian:
\begin{equation}
  H = - J \sum_{<i,j>}\biggl[(1-\Delta)(S_{i}^{x}S_{j}^{x} + S_{i}^{y}S_{j}^{y}) + S_{i}^{z}S_{j}^{z}\biggr]
\end{equation}
where $\mathbf{S}_i=(S_i^{\rm x}, S_i^{\rm y},
S_i^{\rm z})$ is a unit vector in the direction of the
classical magnetic moment at lattice site $i$;
$J>0$ -- ferromagnetic exchange constant;
$\Delta(N)$ characterizes the amount of anisotropy;
$\Delta=0$ corresponds to the isotropic Heisenberg
case; $\Delta=1$ -- the Ising case.
Periodic boundary conditions in the film plane and free boundary conditions in the perpendicular direction were imposed on the system.

The simulations were carried out for systems of size $N_s=L\times L\times N$, where $N$ -- is number of layers and $L=128$ is linear size of layer. We used the Metropolis algorithm for updating spin configurations.
Simulation was carry out at critical temperature
$T_c = 1.15$ for $N = 3$ monolayer (ML),
$T_c = 1.31$ for $N = 5$ ML,
$T_c = 1.39$ for $N = 7$ ML \cite{bib:08,bib:09,bib:10} and different initial state $m_0 = 1$, $m_0=0.0001 \ll 1$.

The effective anisotropy constant $\Delta(N)$ \cite{bib:08,bib:09}
as a function of film thickness $N$ was chosen from experimental studies of the Curie
temperature $T_C$ for thin films of $Ni(111)/W(110)$ \cite{bib:10} with different thicknesses of Ni film.
In this case, the maximum value of the critical temperature corresponds to the maximum value of the anisotropy constant $\Delta(N) = 1$. As the film thickness increases further, the anisotropy constant tends to zero. In this work, the calculations were carried out for $\Delta(N = 3) = 0.636$, $\Delta(N = 5) = 0.734$, $\Delta(N = 7) = 0.816$.

\section{SIMULATION FROM VARIOUS INITIAL STATES}\label{sec_2}

 \begin{figure*}[!t]
	\centering
\includegraphics[width=0.45\textwidth]{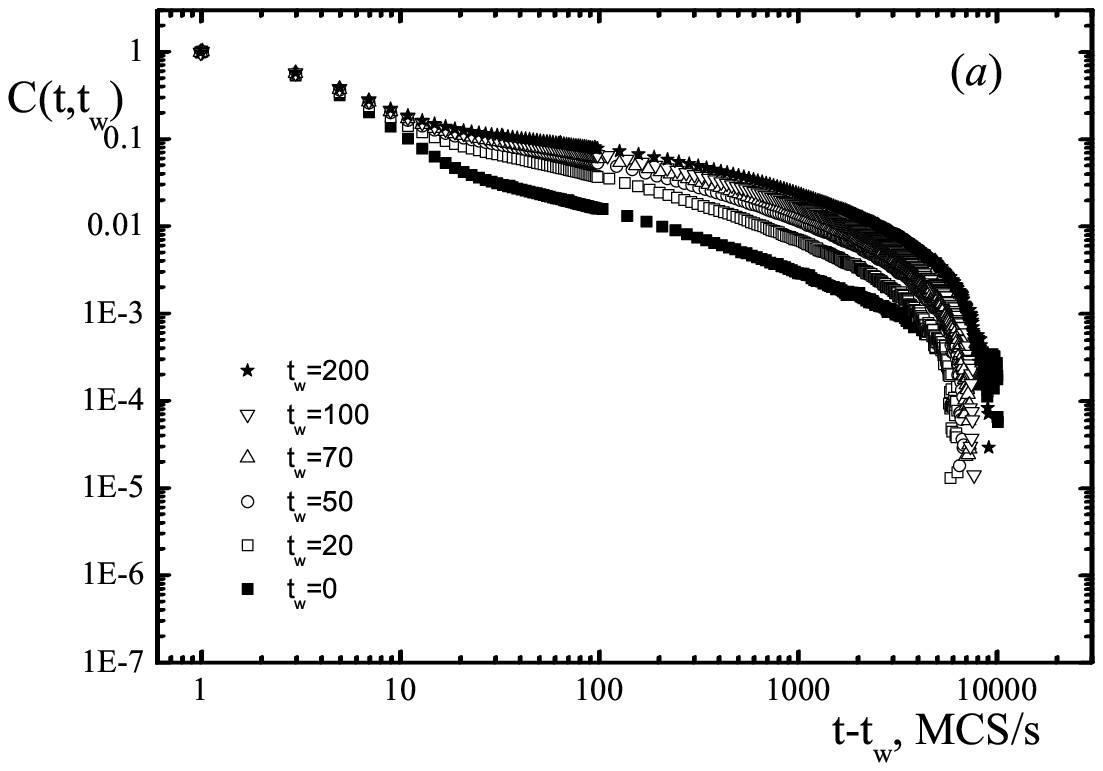}
\includegraphics[width=0.45\textwidth]{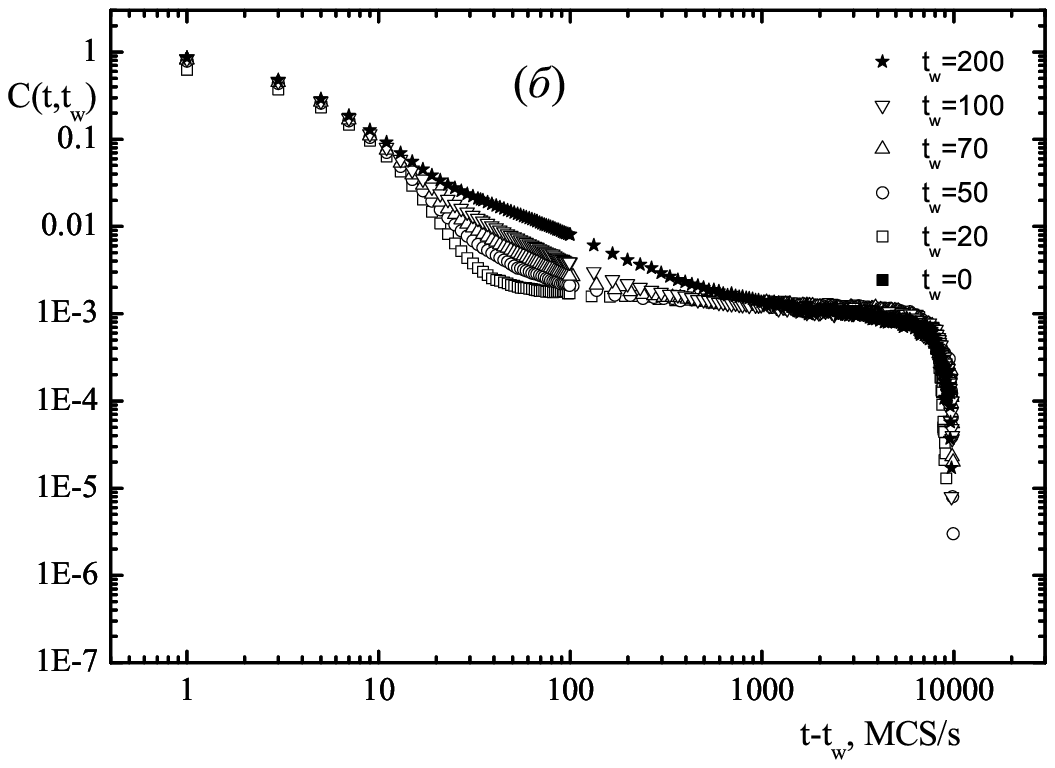}
	\caption{Relaxation of the autocorrelation function $C(t, t_w)$ for a film with a thickness of $N = 3$ ML at different waiting times $t_w = 200$, $100$, $70$, $50$, $20$, $0$ MCS/s from different initial states: high-temperature $m_0\ll 1$ (a) and low-temperature $m_0 = 1$ (b).}
	\label{ris:Corr_N3}
\end{figure*}

We calculated the time dependencies of the two-time autocorrelation function to study the phenomena of aging
\begin{equation}
  C(t,t_{w}) = \left\langle \frac{1}{N_s} \sum_{i}\vec{S}_{i}(t)\vec{S}_{i}(t_{w}) \right\rangle - m(t) \cdot m(t_w) 
\end{equation}
where $t$ is the time from of the sample  preparation; $t_w$ ("waiting time") is time which characterizes the time elapsed since the preparation of the sample prior to measurement of its quantities; $m(t)$ -- magnetization, which is the order parameter for a ferromagnetic film. In Fig. 1, the data for $C(t, t_w)$ are plotted against the observation time $t - t_w$ for a thin Heisenberg film with a thickness $N = 3$ ML for different values of the waiting time $t_w = 200$, $100$, $70$, $50$, $20$, $0$ Monte-Carlo steps per spin (MCS/s).
The simulation was carried out at the critical temperature from the high-temperature initial state $m_0 \ll 1$ (Fig. \ref{ris:Corr_N3}a) and the low-temperature initial state $m_0 = 1$ (Fig. \ref{ris:Corr_N3}b).

The autocorrelation function clearly demonstrates the presence of three characteristic regimes: a quasi-equilibrium regime at time $(t -t_w) \ll t_w$ and a non-equilibrium regime at time $(t - t_w) \gg t_w$. At times $(t - t_w) \sim t_w$, a crossover mode takes place with the correlation characteristics depending on the waiting time \cite{bib:11}. For shorter times $t_w$, the autocorrelation quickly relaxes to a plateau $C_{eq} \sim ~ m^2_{eq} \sim t^{–2\beta/\nu z
}$ and drops to zero only at large values of $t - t_w$. Moreover, for different values of $t_w$, the data are characterized by different laws of decrease, which means that the time invariance is violated. The behavior of the autocorrelation function demonstrates the slowing down of relaxation processes with increasing $t_w$. For example, the autocorrelation function decreases from 1 to 0.01 for $t_w = 20$ after 659 MCS/s, for $t_w = 50$ after 1146 MCS/s, for $t_w = 70$ after 1419 MCS/s, for $t_w = 100$ after 1805 MCS/s, for $t_w = 200$ after 2546 MCS/s when modeling from a high-temperature initial state $m_0 \ll 1$ (Fig. \ref{ris:Corr_N3}a). When modeling from a low-temperature initial state $m_0 = 1$, the relaxation of the autocorrelation function from 1 to 0.01 occurs faster for $t_w = 20$ in 22 MCS/s, for $t_w = 50$ in 26 MCS/s, for $t_w = 70$ in 31 MCS/s, for $t_w = 100$ for 40 MCS/s, for $t_w = 200$ for 78 MCS/s (Fig. \ref{ris:Corr_N3}b).

This violation of time invariance is the second defining property of aging systems together with the slow dynamics mentioned above. Thus, simulation of a dynamic process from a completely ordered state is most preferable due to the lesser influence of fluctuations on the results.

We calculate the dimensionless dynamic correlation function $R(t, t_w)$ \cite{bib:12} to estimate the correlation time of our systems  for different film thicknesses $N = 3$, $5$, $7$ ML and for different waiting times $t_w = 20$, $50$, $70$, $100$, $200$ MCS/s:
\begin{equation}\label{eq:R}
  R(t,t_{w}) = \frac{C(t,t_{w})}{\sqrt{\left[\left\langle\biggl(\frac{1}{N_{s}}\sum_{i}\vec{S}_{i}(t)\vec{S}_{i}(t_{w}) \biggr)^{2}\right\rangle\right]}}\sim e^{-\delta t/\tau_{cor}}.
\end{equation}

\begin{figure*}[!t]
	\centering
\includegraphics[width=0.45\textwidth]{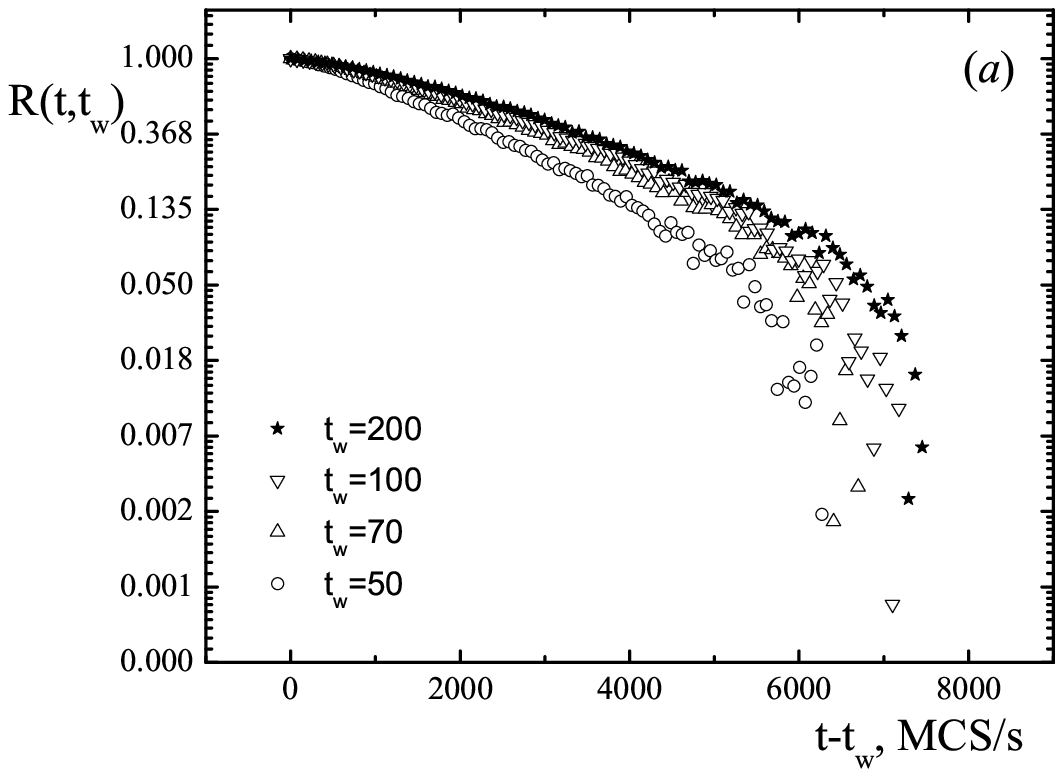}
\includegraphics[width=0.45\textwidth]{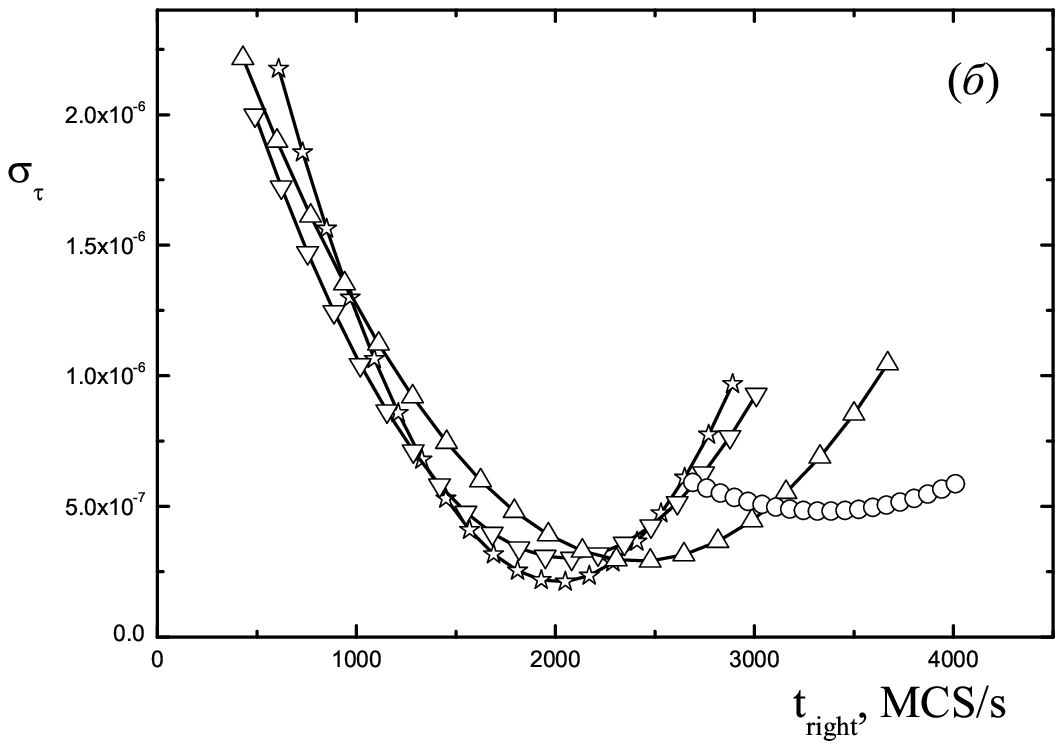}
	\caption{Dimensionless dynamic correlation function $R(t, t_w)$ (a) for thin films with $N = 3$ ML at various waiting times $t_w = 200$, $100$, $70$, $50$, $20$ MCS/s and the corresponding mean-square approximation error (b) from the chosen time interval.}
	\label{ris:R_sigma_N3}
\end{figure*}

The time dependencies of the dimensionless dynamic correlation function $R(t, t_w)$ are shown in Fig. \ref{ris:R_sigma_N3}a. For sufficiently long times, $R(t, t_w)$ decreases exponentially:
\begin{equation}
R(t,t_w) \sim exp(-t/\tau_{cor}).
\end{equation}

We estimated the value of the correlation time $\tau_{cor}$ from the slope of the time dependence of the dimensionless dynamic correlation function $R(t, t_w)$ (Fig. \ref{ris:R_sigma_N3}a) plotted on a logarithmic scale. The minimum mean-square approximation error (Fig. \ref{ris:R_sigma_N3}b) for $N = 3$ is reached in the interval $[700; t_{right}]$ for $t_{right} = 2000$ at $t_w = 200$, for $t_{right} = 2000$ at $t_w = 100$, for $t_{right} = 2400$ at $t_w = 70$, and for $t_{right} = 3500$ at $t_w = 50$.

\begin{figure}[!t]
	\centering
\includegraphics[width=0.45\textwidth]{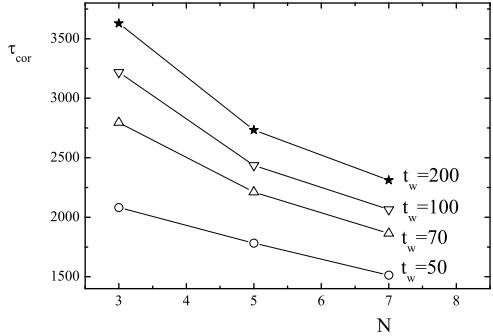}\\
	\caption{Dependence of correlation time on film thickness and waiting time.}
	\label{ris:tau}
\end{figure}

The values of the correlation time are shown in Fig.~\ref{ris:tau}. The value of the correlation time $t_{cor}$ demonstrates the presence of aging effects in thin Heisenberg films. An increase in the age of the system $t_w$ leads to an increase in the value of $\tau_{cor}$.

In spin systems, during the second order phase transitions, critical deceleration effects appear, i.e. an increase in the correlation time $\tau_{cor}$ when approaching the critical point $T_c$. The power character of the asymptotic dependence $\tau_{cor}$ is determined by the universal dynamic critical exponent $z$:
\begin{equation}
\tau_{cor}\sim |T-T_c|^{-\nu z},
\end{equation}
where $\nu$ is the critical exponent of the correlation
length.

For an independent assessment of the dynamic critical exponent $z$, the cumulant $F_2(t)$ was calculated in this work:
\begin{equation}
F_2(t) = \frac{ m^{(2)}(t)|_{m_0=0}}{ m^2(t)|_{m_0=1}} \sim t^{d/z},
\end{equation}
where $d$ is the system dimension.

The time dependence of the cumulant $F_2(t)$ makes it possible to determine the ratio $d/z$ from the slope of the curve plotted on a double logarithmic scale. The following values of this ratio were obtained for different film thicknesses: $d/z = 1.0224(2)$ for $N = 3$ ML, $d/z = 1.0146(3)$ for $N = 5$ ML, $d/z = 1.0985(2)$ for $N = 7$ ML.

The value of the dynamic critical exponents $z$ itself was obtained using the effective system dimension $d_{eff}$, which was obtained from the hyper-scaling relation $\gamma/\nu + 2\beta/\nu = d_{eff}$. Using the values of static critical exponents from [8], the effective dimension of the system was found $d_{eff} = 2.007 (125)$ for films with a thickness of $N = 3$ ML, $d_{eff} = 1.992 (98)$ for films with a thickness of $N = 5$ ML, $d_{eff} = 2.158 (135)$ for films with thickness $N = 7$ ML. The corresponding values of the dynamic critical exponents are $z = 1.827(99)$ for $N = 3$, $z = 1.963(192)$ for $N = 5$, $z = 2.111(131)$ for $N = 7$ were obtained. Thus, films with a thickness of $N = 3$, $N =5$ demonstrate the critical behavior characteristic of quasi-two-dimensional systems.

\section{CONCLUSION}\label{sec_3}

In the critical behavior of thin Heisenberg films, aging effects are manifested. This is indicated by the time behavior of the autocorrelation function. As the waiting time $t_w$ increases, the relaxation processes in the systems slow down. In this work, the correlation time was estimated. An increase in the age of the system $t_w$ leads to an increase in the value of $\tau_{cor}$. The value of the correlation time $\tau_{cor}$ demonstrates the presence of aging effects in thin Heisenberg films.

With a decrease in the size of magnetic systems, fluctuations of the spin density increase and the effects of critical slowing down appear. Thus, aging effects manifest themselves in the non-equilibrium behavior of low-dimensional magnetic systems.

The study was carried out with the financial support of the Ministry of Education and Science of the Russian Federation
(agreement 0741-2020-0002) and grant MD-2229.2020.2 of the President of the Russian Federation.
The computational research was supported in through resources provided by
the Shared Services Center "Data Center of FEB RAS" (Khabarovsk) \cite{bib:13}.

\end{document}